# Electronic structure investigation of $Ti_3AlC_2$, $Ti_3SiC_2$, and $Ti_3GeC_2$ by soft-X-ray emission spectroscopy

M. Magnuson[1], J. -P. Palmquist[2], M. Mattesini[1,4], S. Li[1], R. Ahuja[1], O. Eriksson[1], J. Emmerlich[3], O. Wilhelmsson[2], P. Eklund[3], H. Högberg[3], L. Hultman[3], U. Jansson[2]

[1]*Department of Physics, Uppsala University, P. O. Box 530, S-751 21 Uppsala, Sweden.*

[2]*Department of Materials Chemistry, The Ångström Laboratory, Uppsala University, P.O. Box 538 SE-75121 Uppsala.*

[3]*Department of Physics, IFM, Thin Film Physics Division, Linköping University, SE-58183 Linköping, Sweden.*

[4]*Departamento de Física de la Materia Condensada, Universidad Autónoma de Madrid, E-28049, Spain.*

**Abstract**

The electronic structures of epitaxially grown films of $Ti_3AlC_2$, $Ti_3SiC_2$ and $Ti_3GeC_2$ have been investigated by bulk-sensitive soft X-ray emission spectroscopy. The measured high-resolution Ti *L*, C *K*, Al *L*, Si *L* and Ge *M* emission spectra are compared with *ab initio* density-functional theory including core-to-valence dipole matrix elements. A qualitative agreement between experiment and theory is obtained. A weak covalent Ti-Al bond is manifested by a pronounced shoulder in the Ti *L*-emission of $Ti_3AlC_2$. As Al is replaced with Si or Ge, the shoulder disappears. For the buried Al and Si-layers, strongly hybridized spectral shapes are detected in $Ti_3AlC_2$ and $Ti_3SiC_2$, respectively. As a result of relaxation of the crystal structure and the increased charge-transfer from Ti to C, the Ti-C bonding is strengthened. The differences between the electronic structures are discussed in relation to the bonding in the nanolaminates and the corresponding change of materials properties.

# 1 Introduction

Recently, the interest in nanolaminated ternary $M_{n+1}AX_n$ (denoted 211, 312 and 413, where n=1, 2 and 3, respectively) carbides and nitrides, so-called MAX-phases, has grown significantly. Here, M is an early transition metal, A is a p-element, usually belonging to the groups IIIA and IVA, and X is either carbon or nitrogen [1, 2, 3, 4]. This is due to the fact that these layered materials exhibit a unique combination of metallic and ceramic properties [5], including high strength and stiffness at high temperatures, resistance to oxidation and thermal shock, as well as high electrical and thermal conductivity. These unusual macroscopic properties are closely related to the electronic and structural properties of the constituent atomic layers on the nanoscale.

This family of compounds has a hexagonal structure with near close-packed layers of the M-elements interleaved with square-planar slabs of pure A-elements, where the X-atoms are filling the octahedral sites between the M-atoms. However, the A-elements are also located at the center of trigonal prisms that are larger than the octahedral sites and thus better accomodate the A-atoms. Among the 312 phases, there are three carbides which belong to this family, namely, $Ti_3AlC_2$, $Ti_3SiC_2$ and $Ti_3GeC_2$. Figure 1 shows the crystal structure of the 312 phase.

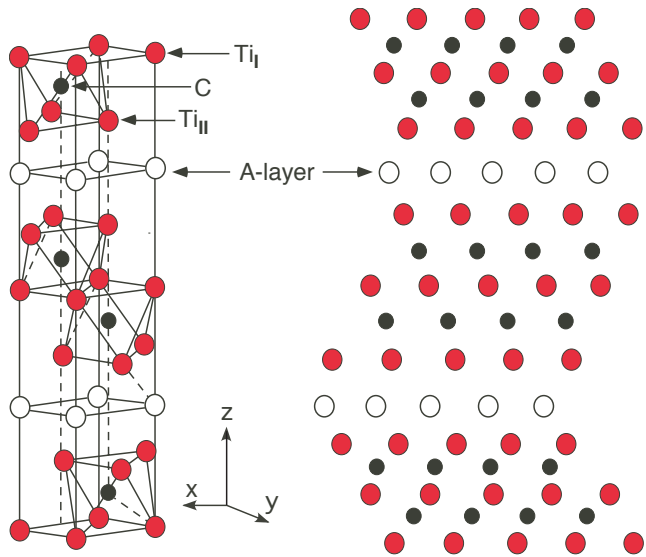

**Figure 1:** The hexagonal crystal structure of the $Ti_3AC_2$-phases. The Ti atoms have two different sites, denoted $Ti_I$ and $Ti_{II}$. Every fourth layer in the 312-phase is interleaved with layers of the pure A-group element.

Up to now, the most studied MAX-material is $Ti_3SiC_2$ which consists of hexagonal layers stacked in the repeated sequence of Si-$Ti_{II}$-C-$Ti_I$-C-$Ti_{II}$, where the unit cell consists of two formula units. Ti atoms can occupy either of two different sites; $Ti_{II}$ which has both C and Si neighbors and $Ti_I$ which only has C neighbors. Most of the research on $Ti_3SiC_2$ has incorporated processing and mechanical properties of sintered bulk compounds [1, 4, 5]. However, in many technological applications where, e.g., high melting points, corrosion resistance, electrical and thermal conductivity as well as low-friction gliding properties are required, high-quality thin film coatings are more useful than bulk materials. Although MAX-phases and related compounds have been studied previously, a full knowledge of why these materials obtain certain properties has not yet been obtained. One reason for why a complete understanding is lacking lies in the difficulties in obtaining accurate electronic structure measurements of internal atomic layers.

Previous experimental investigations of the electronic structure of $Ti_3SiC_2$ include valence-band X-ray photoemission (XPS) [6, 7, 8]. However, XPS is a surface sensitive method which is not element specific. Theoretically, it has been shown by *ab initio* bandstructure calculations that there should be significant differences between the partial density-of-states (DOS) of Ti and C when Si is exchanged for another A-element [9]. Thus, if it would be possible to perform bulk-sensitive and element-specific electronic structure measurements and compare them to *ab initio* bandstructure calculations for a series of compounds with different A-elements, one could obtain the most reliable information about the differences in the internal electronic structure of the buried layers.

In this paper, we investigate and compare the electronic structures of $Ti_3AlC_2$, $Ti_3SiC_2$ and $Ti_3GeC_2$ with each other, using the bulk-sensitive and element-specific soft X-ray emission (SXE) spectroscopy method with selective excitation energies around the Ti 2*p*, C 1*s*, Al 2*p*, Si 2*p* and Ge 3*p* thresholds. The SXE technique is more bulk sensitive than electron-based techniques such as XPS and X-ray absorption spectroscopy (XAS). Due to the involvement of both valence and core levels, each kind of atomic element can be probed separately by tuning the excitation energy to the appropriate core edge. The SXE spectroscopy follows the dipole selection rule and conserves the charge-neutrality of the probed system. This makes it possible to extract both elemental and chemical near ground-state information of the electronic structure. The SXE spectra are interpreted in terms of partial valence band DOS weighted by the transition matrix elements. The main objective of the present investigation is to systematically study the nanolaminated internal electronic structures and the influence of hybridization among the constituent atomic planes in the $Ti_3AC_2$ materials using the combination of X-ray emission spectroscopy and density-functional calculations. By comparing the partial electronic structures of the three 312 systems, important information about the bonding is achieved, creating a basis for the understanding of the unusual materials properties.

# 2 Experimental

## 2.1 X-ray emission measurements

The SXE measurements were performed at the undulator beamline I511-3 at MAX II (MAX-lab National Laboratory, Lund University, Sweden), comprising a 49-pole undulator and a modified SX-700 plane grating monochromator [10]. XAS spectra at the Ti $2p$ and C $1s$ edges were recorded in total electron yield mode by measuring the sample drain current as a function of the photon energy of the incident monochromatized synchrotron radiation. The XAS spectra were normalized to the photo current from a clean gold mesh introduced into the synchrotron radiation beam in order to correct for intensity variations of the incident X-ray beam. During the XAS measurements at the Ti $2p$ and C $1s$ edges, the resolution of the beamline monochromator was about 0.1 eV. The SXE spectra were recorded with a high-resolution Rowland-mount grazing-incidence grating spectrometer [11] with a two-dimensional detector. The Ti $L$ and C $K$ X-ray emission spectra were recorded using a spherical grating with 1200 lines/mm of 5 m radius in the first order of diffraction. The Al $L$, Si $L$ and Ge $M$ spectra were recorded using a grating with 300 lines/mm, 3 m radius in the first order of diffraction. During the SXE measurements at the Ti $2p$, C $1s$, Al $2p$, Si $2p$ and Ge $3p$ edges, the resolutions of the beamline monochromator were 1.6, 1.0, 0.3, 0.2, and 0.4 eV, respectively. The SXE spectra were recorded with spectrometer resolutions 0.7, 0.2, 0.2, 0.2, 0.2 eV, respectively. All the measurements were performed with a base pressure lower than $5\times10^{-9}$ Torr. In order to minimize self-absorption effects [12], the angle of incidence was 30$o$ from the surface plane during the emission measurements. The X-ray photons were detected parallel to the polarization vector of the incoming beam in order to minimize elastic scattering.

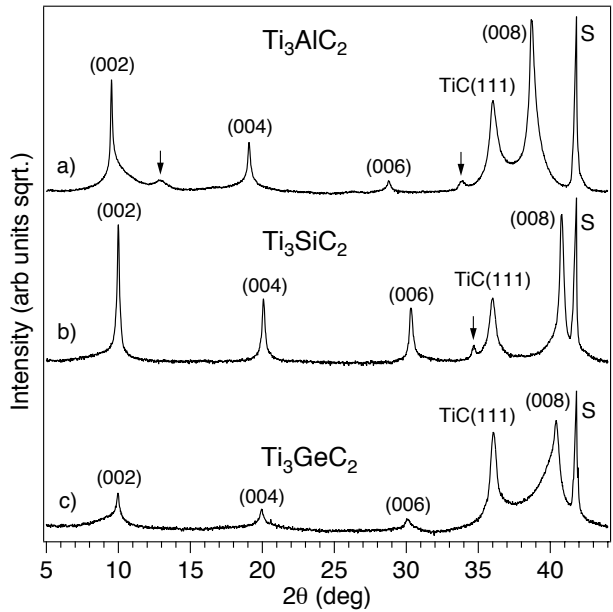

**Figure 2:** X-ray diffraction pattern from $Ti_3AlC_2$, $Ti_3SiC_2$ and $Ti_3GeC_2$ thin films. S denotes the contribution from the substrate. The arrows in a) indicate small contributions of $Ti_2AlC(0002)$ (left) and $Ti_3Al$ (right). The arrow in 2b) corresponds to $Ti_5Si_4C_x$.

## 2.2 Deposition of Ti$_3$AC$_2$ (A= Al, Si, Ge) films

The MAX-phase thin films were deposited by dc magnetron sputtering (base pressure $\sim$ 5x10$^{-10}$ Torr) from elemental targets of Ti and C, and Al, Si or Ge in an argon discharge pressure of 4 mTorr. α-Al$_2$O$_3$ (000*l*) was used as substrate. However, to promote a high quality growth of the MAX-phase, a 200 Å thick seed-layer of TiC$_{0.7}$(111) was initially deposited. The Ti$_3$AlC$_2$ film was 5000 Å thick, while the Ti$_3$SiC$_2$ and Ti$_3$GeC$_2$ films were 2000 Å thick. For further details on the synthesis process, the reader is referred to Refs [13, 14, 15].

Ordinary Θ-2Θ diffractograms of the deposited films are shown in figure 2. As observed, the peaks originate from Ti$_3$AC$_2$(000*l*) together with peaks from the TiC(111) seed-layer and the α-Al$_2$O$_3$(0006) substrate. In figure 2a, small contributions of Ti$_3$Al and Ti$_2$AlC can also be observed, and in figure 2b, a small peak from Ti$_5$Si$_3$C$_x$. (all marked with arrows). The low intensities of the small additional peaks show that the films mainly consist of single-phased MAX-phase. Furthermore, the fact that the diffractograms only show Ti$_3$AC$_2$ of {000*l*}-type suggests highly textured or epitaxial films. X-ray pole figures verified that the growth indeed was epitaxial, and determined the relation to Ti$_3$AC$_2$ (000*l*)//TiC(111)//Al$_2$O$_3$(000*l*) with an in-plane orientation of Ti$_3$AC$_2$[210]//TiC[110]//Al$_2$O$_3$[210]. The epitaxial growth behaviour has also been documented by transmission electron microscopy (TEM) [16, 17, 18, 19, 20]. XPS-analysis depth profiles of the deposited films within the present study using a PHI Quantum instrument, showed after 60 seconds of Ar-sputtering a constant composition without any contamination species. From the diffractograms in figure 2, the values of the c-axis were determined to be 18.59 Å, 17.66 Å and 17.90 Å for the Ti$_3$AlC$_2$, Ti$_3$SiC$_2$ and Ti$_3$GeC$_2$ MAX-phases, respectively.

# 3 Computational details

## 3.1 Calculation of the X-ray emission spectra

The X-ray emission spectra were computed within the single-particle transition model by using the full potential linearized augmented plane wave (FPLAPW) method [21]. Exchange and correlation effects were taken into account through the generalized gradient approximation (GGA) as parameterized by Perdew, Burke and Ernzerhof [22]. A plane wave cut-off, corresponding to $R_{MT}*K_{max}$=8, was used for all the investigated phases. The charge density and potentials were expanded up to $\ell$=12 inside the atomic spheres, and the total energy was converged with respect to the Brillouin zone integration.

The emission intensity $I_c$ for a hole created in the c:th core shell can be written in cgs units as [23]

$$I_c(E) = \frac{2e^2\hbar\omega^3}{mc^3} F_c(E), \quad E<E_F \qquad (1)$$

where $F_c(E)$ represents the spectral distribution given by [23, 24, 25]

$$F_c(E) = \frac{2m}{3\hbar^2}(\hbar\omega) \sum_{\vec{k}j} \sum_{M=-J}^{J} |\langle \varphi_{cM}|\hat{\varepsilon}_{\vec{q}} \cdot \vec{r}|\psi_{\vec{k}j}\rangle|^2 \delta(E-E_{\vec{k}j}). \qquad (2)$$

In the above equation $\varphi_{cM}$ is the core wave function, $\vec{q}$ the wave vector of the incident photon, $\hat{\varepsilon}_q$ the polarization tensor, and $E_{\vec{k}j}$ and $\psi_{\vec{k}j}$ are the energy and the wave function of the $j$th valence band at vector $\vec{k}$. The energy of the emitted photon is represented by $\omega = E - E_c$, where $E_c$ is the core energy level.

The emission spectra were computed by using the electric-dipole approximation which means that only the transitions between the core states with orbital angular momentum $\ell$ to the $\ell \pm 1$ components of the valence bands were considered. The core-hole lifetimes used in the calculations were 0.73 eV, 0.27 eV, 0.45 eV, 0.5 eV and 2.0 eV, for the Ti $2p$, C $1s$, Si $2p$, Al $2p$ and Ge $3p$ edges, respectively. A direct comparison of the calculated spectra with the measured data was finally achieved by including the instrumental broadening in form of Gaussian functions corresponding to the experimental resolutions (see experimental section IIA). The final state lifetime broadening was accounted for by a convolution with an energy-dependent Lorentzian function with a broadening increasing linearly with the distance from the Fermi level ($E_F$) according to the function $a+b(E-E_F)$, where the constant $a$ is in units of eV and $b$ is dimensionless [26]. For Ti, C, Al and Si $a$ was 0.01 eV and $b$ was 0.05. For Ge, a variable lifetime broadening of the valence band was not feasible since the expanded energy region also contains the $3d$ core-levels.

## 3.2 Balanced crystal orbital overlap population (BCOOP)

In order to study the chemical bonding of the $Ti_3AlC_2$, $Ti_3SiC_2$ and $Ti_3GeC_2$ compounds, we also calculated the BCOOP function by using the full potential linear muffin-tin orbital (FPLMTO) method [27]. In these calculations, the muffin tins are kept as large as possible (without overlapping one another), so that the muffin tins fill about 66% of the total volume. To ensure a well-converged basis set, a double basis with a total of four different $\kappa^2$ values is used. For Ti, we include the $4s$, $4p$ and $3d$ as valence states. To reduce the core leakage at the sphere boundary, we also treat the $3s$ and $3p$ core states as semi-core states. For Al and Si, $3s$, $3p$ and $3d$ are taken as valence states. For Ge, we used semi-core $3d$ and valence $4s$, $4p$ and $4d$ basis functions. The resulting basis forms a single, fully hybridizing basis set. This approach has previously proven to give a well-converged basis [28]. For the sampling of the irreducible wedge of the Brillouin zone, we use a special-k-point method [29] and the number of k points we used is 216 for the self-consistent total energy calculation. In order to speed up the convergence, a Gaussian broadening of width 20 mRy is associated with each calculated eigenvalue.

## 4 Results

### 4.1 Ti $L$ X-ray emission spectra

Figure 3 (top) shows Ti $L_{2,3}$ SXE spectra of $Ti_3AlC_2$, $Ti_3SiC_2$, and $Ti_3GeC_2$ excited at 458 eV, 459.8 eV, 463.5 eV and 477 eV photon energies, corresponding to the $2p_{3/2}$ and $2p_{1/2}$ absorption maxima (vertical bars in the XAS spectra) and nonresonant excitation. The XAS measurements have been made to identify the resonant excitation energies for the SXE measurements. For comparison of the SXE spectral profiles, the measured spectra are normalized to unity and are plotted on a common photon energy scale and relative to the $E_F$ using the $2p_{1/2}$ core-level XPS binding energy of 460.8 eV in $Ti_3SiC_2$ [8]. The main $L_3$ and $L_2$ emission lines are observed at -8.7 eV and -2.5 eV on the relative energy scale. Contrary to pure Ti, a peak structure is also observed at -16.2 eV. This is the result of Ti - C hybridization. As observed, the Ti $L_{2,3}$ SXE spectra appear rather delocalized (wide bands) which makes electronic structure calculations suitable for the interpretation of the spectra. Calculated spectra for the two Ti sites ($Ti_I$ and Ti$_{II}$), their sum and difference are also shown in the lower part of Figure 3. Although SXE is a site-selective spectroscopy, separate contributions from the two crystallographically different Ti sites are not experimentally resolved due to their negligible energy difference. For resonant excitation at the $2p_{1/2}$ absorption maximum (463.5 eV), the $L_2$ emission resonates and the $L_3/L_2$ ratio is close to 2:1. For nonresonant, continuum excitation far above the $2p$ thresholds (477 eV), the $L_3/L_2$ ratio is 6:1. In general, nonresonant $L_3/L_2$ emission ratios in $3d$ transition metals have been observed to be higher than the statistical weight of atomic levels (2:1) due to the opening of the Coster-Kronig process at the $2p_{1/2}$ threshold. The Coster-Kronig decay from the $2p_{1/2}$ core-level leads to a shorter lifetime and a larger Lorentzian width for the $2p_{1/2}$ core state than for the $2p_{3/2}$ state. For a clear comparison between the experimental and calculated spectral structures, the nonresonant emission spectra excited at 477 eV are suitable since they resemble the occupied electronic states, and resonant phenomena are avoided. In the fitting procedure to the 477 eV SXE spectra, we employed the experimental values for the $L_3/L_2$ ratio of 6:1 and the $L_{2,3}$ peak splitting of 6.2 eV which is larger than our calculated *ab initio* spin-orbit splitting of 5.7 eV. Note that the Ti $2p$ peak splitting obtained from XAS, XPS [30] and SXE [31] spectra are not exactly the same due to screening and different number of electrons in the final states. The spectral weight at the E$_F$ is significantly higher for the calculated Ti$_{II}$ atoms which are directly bonded to the A atoms than for the Ti$_I$ atoms. This suggests that the Ti$_{II}$ atoms contribute more to the conductivity than the $Ti_I$ atoms. The $Ti_{II}$-$Ti_I$ difference spectra are a measure of the difference in the electronic structure and hence, the bonding strength between the 3*d* states of the $Ti_{II}$ and $Ti_I$ atoms and the *s,p,d* states of the A-element. The largest difference in the $Ti_{II}$ - $Ti_I$ bonding is observed for the $Ti_3AlC_2$ system. The most significant difference between the studied systems is the pronounced shoulder (observed both in experiment and calculations) in $Ti_3AlC_2$ (indicated by the arrow at the top in Fig. 3) which is not observed in the other systems. This shoulder has a splitting from the main line of 1.5 eV. From the calculated bandstructure, the nature of this shoulder is related to a series of flat bands (not shown) with Ti 3*d* character which are shifted towards the $E_F$.

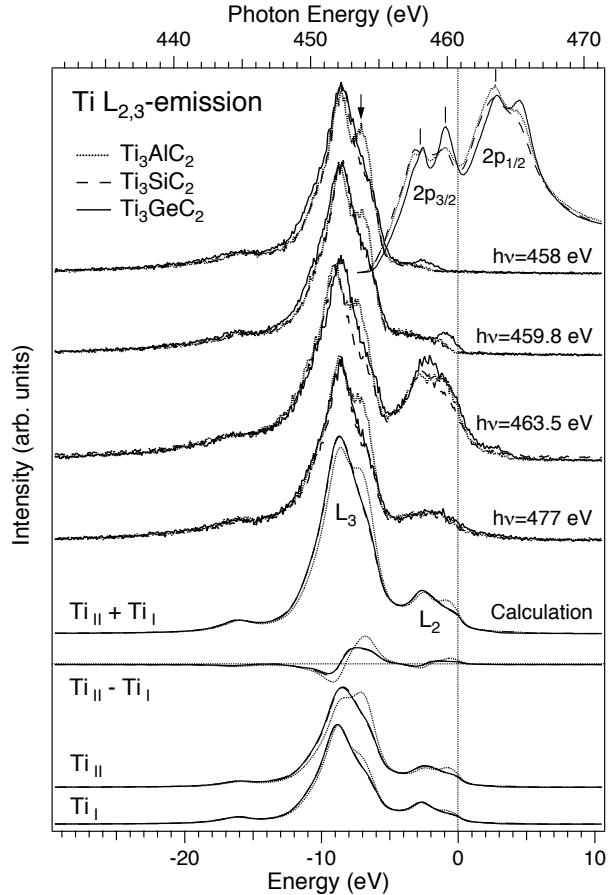

**Figure 3:** Top: experimental Ti $L_{2,3}$ SXE spectra of Ti$_3$AlC$_2$, Ti$_3$SiC$_2$ and Ti$_3$GeC$_2$ excited at 458, 459.8, 463.5 and 477 eV, indicated by the vertical bars in the XAS spectra. The XAS spectra are aligned to the $2p_{1/2}$ threshold and the $2p_{3/2,1/2}$ peak splitting is indicated. Bottom: fitted spectra of the two different Ti sites, Ti$_I$ and Ti$_{II}$ using the experimental $L_{2,3}$ peak splitting of 6.2 eV and the $L_3/L_2$ ratio of 6:1 in comparison to the 477 eV experimental spectra. Note that the fitted spectra for Ti$_3$SiC$_2$ and Ti$_3$GeC$_2$ are almost identical. The sum and difference of the Ti$_{II}$ and Ti$_I$ contributions are shown in the middle. The arrow indicates the shoulder of the Ti$_3$AlC$_2$ system and the E$_F$ is indicated by the vertical dotted line.

## 4.2 C *K* X-ray emission spectra

Figure 4 (top) shows experimental C *K* SXE spectra of $Ti_3AlC_2$, $Ti_3SiC_2$, and $Ti_3GeC_2$ excited at 284.8 eV and 310 eV photon energies. The XAS measurements have been made to identify the resonant excitation energy (vertical bar) for the SXE measurements. The experimental spectra are plotted both on a photon energy scale and relative to the $E_F$ using the C 1*s* core-level XPS binding energy of 281.83 eV in $Ti_3SiC_2$ [8]. Calculated spectra are shown at the bottom of Figure 4. The main peak at -2.6 eV has shoulders on both the low and high-energy sides at -4.2 eV and -2 eV. The agreement between the experimental and calculated spectra is good although the low-energy shoulder at -4.2 eV is more pronounced in the experiment. The main peak and the shoulders correspond to the occupied C 2*p* orbitals hybridized with the Ti 3*d* and A *spd* bonding and antibonding orbitals of the valence bands. As observed both in the experimental and calculated spectra, the high-energy shoulder is less pronounced in $Ti_3AlC_2$ in comparison to $Ti_3SiC_2$ and $Ti_3GeC_2$ which have similar spectral shapes. It should also be noted that the chemical shift between $Ti_3AlC_2$ and the other two studied compounds is very small both in the experiment and in the theoretical work. Experimentally, the upward chemical shifts for $Ti_3AlC_2$ and $Ti_3SiC_2$ are both ~ 0.05 eV while the theory predicts a somewhat larger upward shift of +0.24 eV for $Ti_3AlC_2$, in contrast to observations.

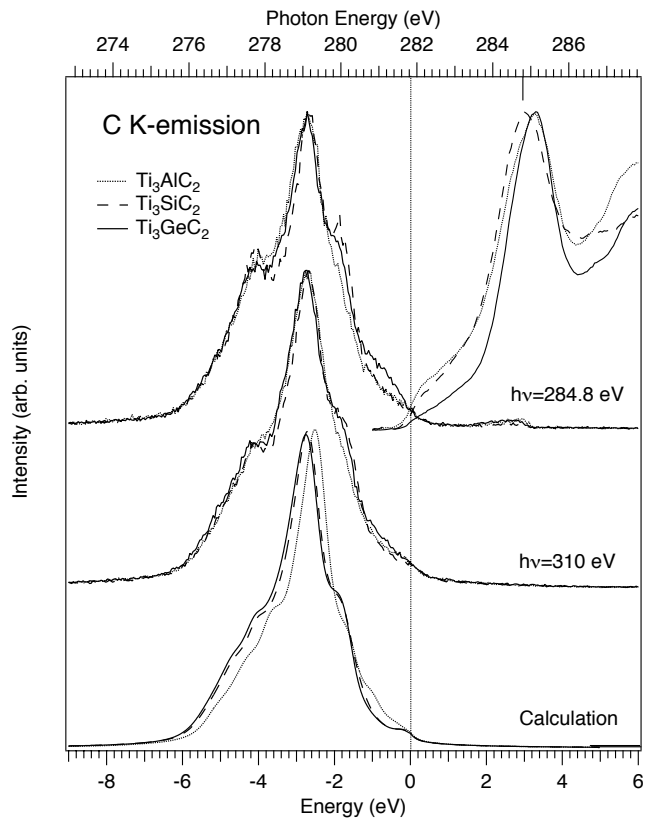

**Figure 4:** Top: Experimental C *K* SXE spectra excited at 284.8 and 310 eV. The resonant excitation energy of 285 eV is indicated by the vertical bar in the XAS spectra. Bottom: calculated spectra. The vertical dotted line indicates the $E_F$.

### 4.3 Al *L* X-ray emission spectra

Figure 5 (top) shows an experimental Al $L_{2,3}$ SXE spectrum of $Ti_3AlC_2$ measured nonresonantly at 120 eV photon energy. The experimental spectrum is plotted on a photon energy scale and relative to the $E_F$ using the Al $2p_{1/2}$ core-level XPS binding energy of 71.9 eV for $Ti_3AlC_2$ [7]. A calculated spectrum with the $L_{2,3}$ spin-orbit splitting of 0.438 eV and the $L_3/L_2$ ratio of 2:1 is shown at the bottom. Comparing the experimental and calculated spectra, it is clear that the main peak at -3.9 eV of the SXE spectrum is dominated by 3*s* final states. The partly populated 3*d* states form the broad peak structure close to the Fermi level and participate in the Ti-Al bonding in $Ti_3AlC_2$. Al 3*p* states dominate in the upper part of the valence band but do not directly contribute to the $L_{2,3}$ spectral shape since they are dipole forbidden. The contribution of the Al 3*p* states can be probed using SXE at the Al *K*-edge. For the Al $L_{2,3}$ SXE spectrum, the valence-to-core matrix elements are found to play an important role to the spectral shape. In contrast to Al $L_{2,3}$ SXE spectra of pure Al, which have a sharp and dominating peak structure within 1 eV below $E_F$, the Al $L_{2,3}$ SXE spectrum of $Ti_3AlC_2$ has a strongly modified spectral weight towards lower energy. A similar modification of the Al $L_{2,3}$ SXE spectral shape has also been observed in the metal aluminides [32]. Comparing the spectral shape to the aluminides, the appearance of the broad low-energy shoulder around -6 eV in the Al $L_{2,3}$ SXE spectrum of $Ti_3AlC_2$ can be attributed to the formation of hybridized Al 3*s* states produced by the overlap of the Ti 3*d*-orbitals. This interpretation is supported by our first principle calculations.

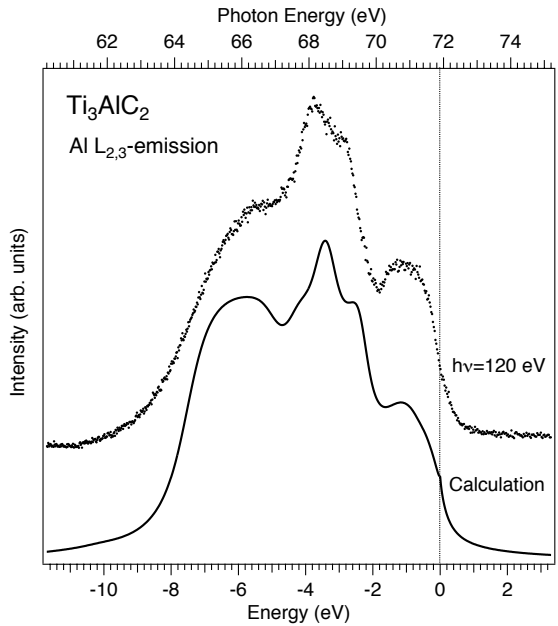

**Figure 5:** Top: Experimental Al $L_{2,3}$ SXE spectrum of $Ti_3AlC_2$. A calculated spectrum is shown below. The vertical dotted line indicates the $E_F$.

### 4.4 Si *L* X-ray emission spectra

Figure 6 (top) shows an experimental Si $L_{2,3}$ SXE spectrum of $Ti_3SiC_2$ measured nonresonantly at 120 eV photon energy. The experimental spectrum is plotted on a photon energy scale and relative to the $E_F$ using the Si $2p_{1/2}$ core-level XPS binding energy of 99.52 eV for $Ti_3SiC_2$ [8]. A calculated spectrum with the $L_{2,3}$ spin-orbit splitting of 0.646 eV and the $L_3/L_2$ ratio of 2:1 is shown at the bottom. In the Si spectrum, the 3*s*-character is concentrated to the bottom of the valence band with the main peak at -7 eV. The shift of the main peak towards lower energy is related to the extra valence electron in Si in comparison to Al.

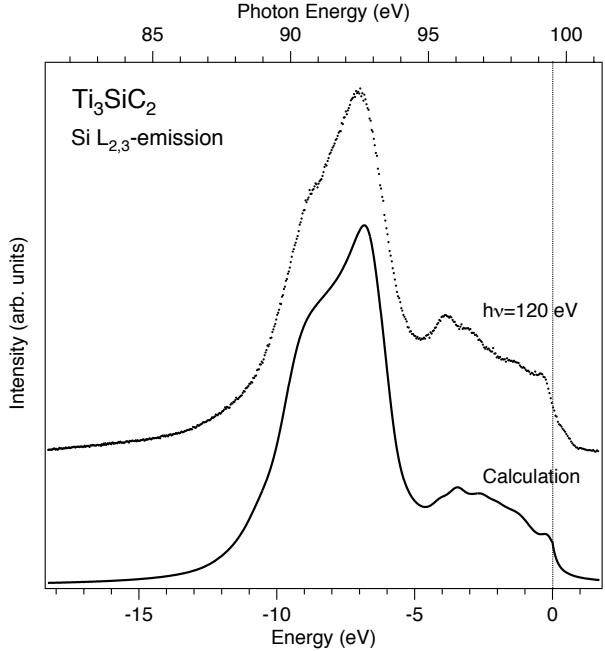

**Figure 6:** Top: Experimental Si $L_{2,3}$ SXE spectrum of Ti$_3$SiC$_2$. A calculated spectrum is shown below. The vertical dotted line indicates the $E_F$.

As in the case of Al, the partly populated Si 3$d$ states in the upper part of the valence band also participate in the Ti-Si bonding in Ti$_3$SiC$_2$. The Si 3$p$-states in the upper part of the valence band do not directly contribute to the $L_{2,3}$ spectral shape since they are dipole forbidden, but can be probed using SXE at the Si $K$-edge. The valence-to-core matrix elements are found to play an important role to the spectral shape of the Si $L_{2,3}$ SXE spectrum. Notably, the observed spectral shape of the Si $L_{2,3}$ SXE spectrum of Ti$_3$SiC$_2$ appears rather different from single crystal bulk Si which has a pronounced double-structure [33, 34]. On the other hand, the spectral shape of the Si $L_{2,3}$ SXE spectrum of Ti$_3$SiC$_2$ is similar to what has been observed for the metal silicides [35, 36]. Comparing the Si $L_{2,3}$ SXE spectrum of Ti$_3$SiC$_2$ to the silicides, the low-energy shoulder on the main peak around -8.5 eV can be attributed to the formation of hybridized Si 3$s$ states produced by the overlap of the Ti 3$d$-states, which is also supported by our theory.

## 4.5 Ge $M$ X-ray emission spectra

Figure 7 (top) shows experimental Ge $M_{2,3}$ SXE spectra of Ti$_3$GeC$_2$ measured nonresonantly at 150 eV (27 eV above the 3$p_{3/2}$ absorption maximum) and 165 eV photon energies. The experimental spectra are plotted on a photon energy scale and relative to the $E_F$ using the Ge 3$p_{1/2}$ core-level XPS binding energy of 125.5 eV for Ti$_3$GeC$_2$ [37]. A calculated spectrum is shown at the bottom, both for the 4$sd$ valence band (full curve) and the 3$d$ core levels (dashed curve). It should be noted that the measured Ge $M_{2,3}$ emission is two orders of magnitude weaker than the Al and Si $L_{2,3}$ emission which makes the measurements more demanding. The shallow 3$d$ core levels shown at the left consists of two peaks from the Ge $M_{4,5} \rightarrow M_3$ and $M_4 \rightarrow M_2$ (3$d \rightarrow$3$p_{3/2,1/2}$) transitions with energies of -32.7 and -29.1 eV relative to the $E_F$. The observed Ge $M_{2,3}$ spin-orbit splitting of 3.6 eV [38] is somewhat smaller than the calculated value of 4.3 eV and the calculated 3$d$ core levels (dashed curve) are also closer to the $E_F$ by 3.9 eV. The measured $M_3/M_2$ intensity ratio of about 6/5:1 differs from the statistical ratio of 2:1. The intensity of the 3$d$ core-levels was divided by a factor of 10 to match the experimental data. In addition to the calculated DOS, the overlap by the matrix elements play an important role for the intensity of the shallow 3*d* core levels. However, the general agreement between the measured and calculated spectral shapes is good.

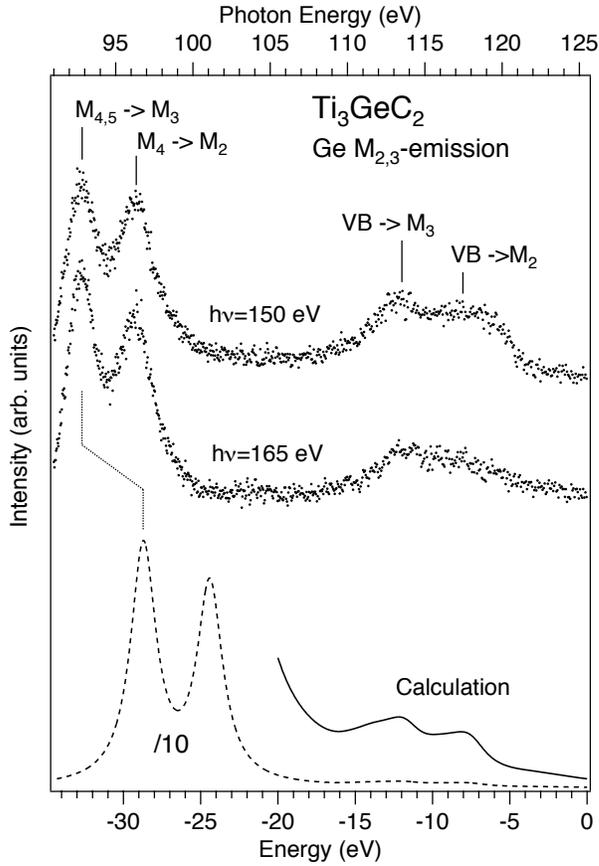

**Figure 7:** Top: Experimental Ge $M_{2,3}$ SXE spectra of $Ti_3GeC_2$ excited at 150 and 165 eV. A calculated spectrum is shown below. For comparison, the calculated spectrum has been divided by a factor of 10 for the 3*d* core level.

The 4*sd* valence band within 10 eV from the $E_F$ consists of a double structure with $M_3$ and $M_2$ emission peaks observed at -12 eV and -8 eV, respectively. This double structure is significantly different from the three-peak structure observed in the $M_{2,3}$ emission of bulk Ge [39]. In bulk Ge, the 4*sd* DOS of the $M_3$ and $M_2$ emission bands are not only separated by the spin-orbit splitting of 3.6 eV, but are also split up into two subbands, separated by 3.5 eV (as in the case of single crystal bulk Si). The result is a triple structure in the valence band of bulk Ge where the upper and lower emission peaks are solely due to $M_3$ and $M_2$ emissions, respectively, while the main middle peak is a superposition of both $M_3$ and $M_2$ contributions. On the contrary, the Ge 4*sd* DOS in $Ti_3GeC_2$ consists of a single peak structure with more spectral weight towards the $E_F$. A shift of peak structures towards the $E_F$ in comparison to bulk Ge has been observed in Ge *K* emission (probing the 4*p* valence states) of the metal-germanides [40]. The Ge 4*p*-states in the upper part of the valence band are not directly observed in the $M_{2,3}$ emission spectra since they are dipole forbidden but are indirectly reflected due to the mixing of *s* and *p* states in the solid. Finally, we note that at 150 eV excitation energy, a weak energy loss structure is superimposed onto the valence band. This weak loss structure, tracking 30 eV below the excitation energy, originates from localized *dd* excitations from the shallow 3*d* core level. At the excitation energy of 165 eV (42 eV above the $3p_{3/2}$ absorption maximum) the dispersing loss structure does not affect the spectral profile of the valence band.

## 4.6 Chemical Bonding

By relaxing the cell parameters of the $Ti_3AC_2$ phases (A=Al, Si and Ge), it was possible to calculate the equilibrium c-axis. They were determined to be 18.64 Å, 17.68 Å and 17.84 Å for $Ti_3AlC_2$, $Ti_3SiC_2$ and $Ti_3GeC_2$, respectively. These values are in excellent agreement with the experimental values in section II B. In order to analyze the chemical bonding, we show in Figure 8 the calculated BCOOP of the three $Ti_3AC_2$ systems [41, 42]. This analysis provides more information about the bonding and is the partial DOS weighted by the balanced overlap population. The BCOOP is a function which is positive for bonding states and is negative for anti-bonding states. The strength of the covalent bonding can be determined by summing up the area under the BCOOP curve. The energy position of the peaks also gives an indication of the strength of the covalent bonding. Firstly, comparing the areas under the BCOOP curves and the distances of the main peaks around -3 eV from the $E_F$, it is clear that the Ti $3d$ - C $2p$ bonds are generally much stronger than the Ti $3d$ - A-element $p$ bonds. Additional bonding occurs around -10 eV with an important Ti $3d$ - C $2s$ overlap, corresponding to -16.2 eV when the $L_{2,3}$ peak splitting is taken into account (see Fig. 3). The areas under the $Ti_{II}$-C curves are larger than for the $Ti_I$-C curves, which indicates stronger bonds. This implies that the $Ti_{II}$ atoms loose some bond strength to the nearest neighbour A atoms, which to some degree is compensated with a stronger $Ti_{II}$-C bond. This observation implies that the Ti-C bonds in principle can be stronger in the $Ti_3AC_2$-phase than in TiC. This can also be observed in our calculated value for the Ti-C bond length which is 2.164 Å for TiC, in good agreement with the experimental value of 2.165 Å [4]. For comparison, the calculated shorter $Ti_{II}$ - C bond lengths for $Ti_3AlC_2$, $Ti_3SiC_2$ and $Ti_3GeC_2$ are 2.086 Å, 2.097 Å and 2.108 Å, respectively, indicating stronger bonds. Secondly, comparing the BCOOP curves between the three systems show that the $Ti_{I,II}$-C BCOOP curves of $Ti_3AlC_2$ are the most intense and are somewhat shifted towards the $E_F$. Thirdly, the $Ti_{II}$ $3d$ - Al $3p$ BCOOP peak is located at about -1 eV below $E_F$ (indicated by the arrow), while the Ti$_{II}$ 3$d$ - Si 3$p$ and Ti$_{II}$ 3$d$ - Ge 4$p$ BCOOP peaks are located around -2 eV below the E$_F$. This is an indication that the Ti$_{II}$ 3$d$ - Al 3$p$ bond is weaker than the Ti$_{II}$ 3$d$ - Si 3$p$ and Ti$_{II}$ 3$d$ - Ge 4$p$ bonds which has also been confirmed by the differences in bond lengths [9]. Our calculated Ti$_{II}$ - A bond lengths for Ti$_3$AlC$_2$, Ti$_3$SiC$_2$ and Ti$_3$GeC$_2$ are 2.885 Å, 2.694 Å and 2.763 Å, respectively. Fourthly, the calculated C 2$p$ - A $s$ overlaps (not shown) have significantly different shapes in comparison to the other overlaps which is an indication that this bond has a weaker covalent character.

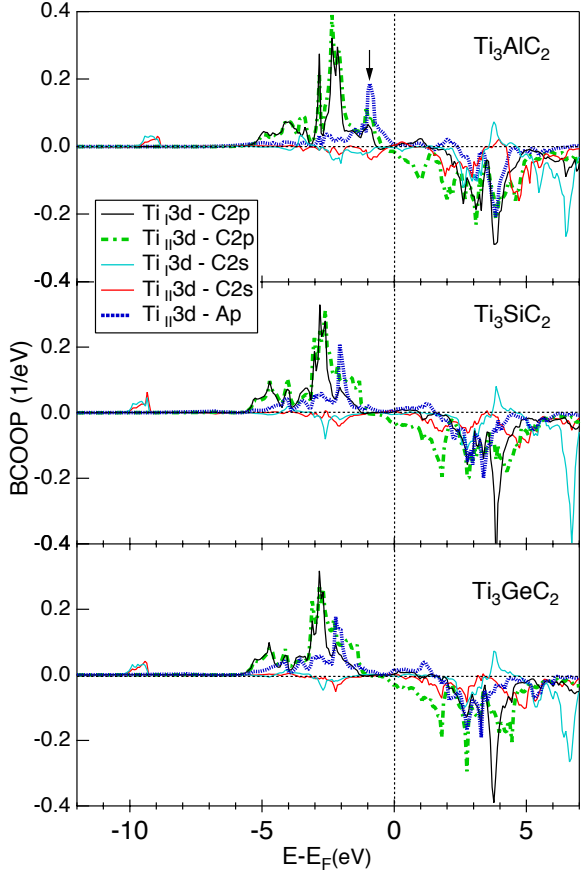

**Figure 8:** Calculated balanced crystal overlap population (BCOOP) for $Ti_3AlC_2$, $Ti_3SiC_2$ and $Ti_3GeC_2$. The arrow indicates the energy shift of the $Ti_{II}$ $3d$-Al $3p$ orbital overlap.

For the Al $L_{2,3}$ SXE spectrum of Ti$_3$AlC$_2$, presented in section C, the BCOOP calculations confirm that the broad low energy shoulder at -6 eV can be attributed to bonding Al 3$s$ orbitals hybridized with Ti 3$d$ and C 2$p$ orbitals. The main peak at -3.9 eV corresponds to hybridization with bonding C 2$p$ orbitals and the shoulder at -3 eV is due to both bonding Ti 3$d$ and C 2$p$ orbitals. The broad structure around -1 eV is due to bonding Ti 3$d$ orbitals and antibonding C 2$p$ orbitals. In the case of the Si $L_{2,3}$ SXE spectrum of Ti$_3$SiC$_2$ presented in section D, the shoulder at -8.5 eV is attributed to hybridization with bonding Ti 3$d$ orbitals and antibonding C 2$p$ orbitals. The peak structure at -3.5 eV corresponds to hybridization with bonding Ti 3$d$ orbitals and antibonding C 2$p$ orbitals while the wiggles between -1 eV and -3 eV are related to bonding Ti 3$d$ and C 2$p$ orbitals of varying strength.

# 5 Discussion

The present results provide new understanding of the electronic structure and hence the bonding in the Ti$_3$AC$_2$-phases (A=Al, Si, Ge) as a basis for an understanding of their unusual materials properties. Our SXE investigation confirms that the Ti 3$d$ - A $p$ bonding is different in Ti$_3$AlC$_2$ in comparison to Ti$_3$SiC$_2$ and Ti$_3$GeC$_2$. The calculations show that the double peak in the Ti $L_{2,3}$ SXE spectrum mainly originate from the Ti$_{II}$ atoms. As a consequence of increasing the c-axis of the hexagonal 312 unit cell to its equilibrium length (i.e., increasing the c-distance in the z-direction in Fig. 1) and the simultaneous symmetry breaking of the A-element mirror plane in the trigonal prisms, the Ti$_{II}$ 3$d$ - A $p$ bond is shortened and the Ti$_I$ 3$d$ - C $p$ bond is lengthened. In the case of Ti$_3$SiC$_2$ and Ti$_3$GeC$_2$, the splitting of the double-structure stemming from the Ti$_{II}$ atoms in the calculated Ti 3$d$ DOS is considerably smaller ($\sim$ 0.5 eV) than for Ti$_3$AlC$_2$ (1.5 eV). As a consequence, no double-peak structure is resolved and observed in the sum of the Ti$_I$ and Ti$_{II}$ peak contributions for Ti$_3$SiC$_2$ and Ti$_3$GeC$_2$. The significantly larger peak splitting in Ti$_3$AlC$_2$ is related to the difference in bonding character as shown by the BCOOP analysis in Fig. 8. This implies that the Ti$_{II}$ 3$d$ - Al 3$p$ bond is less covalent than the Ti$_{II}$ 3$d$ - Si 3$p$ and Ti$_{II}$ 3$d$ - Ge 4$p$ bonds. Our results confirm the general trend in the Ti-A bond strength where the Ti-Si and Ti-Ge bonds are stronger than the Ti-Al bond [9]. As the valence electron population of the A-element is increased by replacing the partly filled valence band of a IIIA element by the isoelectronic valence band of a IVA element e.g., by replacing Al to Si or Ge, more charge is placed into the Ti-A bond. This is directly reflected in a stronger bond and implies a change in the elastic properties. Ti$_3$SiC$_2$ and Ti$_3$GeC$_2$ are therefore stiffer and have significantly larger Young's modulus and electrical conductivity than Ti$_3$AlC$_2$. However, all MAX-phases are generally softer and have lower Young's modulus and higher electrical conductivity than the parent compound TiC. This makes crystallographically oriented MAX-phase films particularly useful as coating material in applications where e.g., high temperatures, electrical and thermal conductivity, and low-friction gliding/sliding properties are required. One example of an application where MAX-phase materials have been shown to be useful is components such as electrical switches and contacts.

An interesting result is that the $Ti_{II}$-C bond is stronger in the MAX-phase than in TiC, as suggested by the differences in bond lengths discussed in Section F. This may seem contradictory but can be understood from the following arguments. If the A-atoms in the MAX-phase are replaced by C, the result is a highly twinned $Ti_3C_3$ structure (see Fig. 1). The result of de-twinning is monocarbide TiC. Preliminary calculations (not shown) suggest that a replacement of all A-elements by C in the 312-crystal structure results in a substantially smaller charge-transfer from Ti to C compared to the MAX-phase. Equivalently, inserting A-element planes into a highly twinned TiC structure implies a substantial charge-transfer from Ti to C and a strengthening of the $Ti_{II}$-C bond. Here, the weaker $Ti_{II}$-A bond in one direction is compensated by the strengthening of the $Ti_{II}$-C bond in another direction. The BCOOP analysis confirms that the covalent Ti $3d$ - C $2p$ bond is generally much stronger than the Ti $3d$-A $p$ bond in the MAX-phase. The same type of observation has been made when point vacancies are introduced into monocarbide, TiC [43]. The insertion of vacancies into TiC resembles that A-planes are introduced and implies a strengthening of the Ti-C bond. This is the major reason for the nanolaminated character in the MAX-phases with delamination by the Ti-A bond breaking as an important step in the deformation mechanism [4]. It is also known that a larger number of the same kind of bonds in a system reduces the strength of each individual bond [41, 44]. The influence of different A-elements on the bond strength in the MAX-phases will be subject to further investigation.

# 6 Conclusions

We have made the first systematic electronic structure investigation of $Ti_3AlC_2$, $Ti_3SiC_2$ and $Ti_3GeC_2$ using the combination of soft X-ray emission spectroscopy and calculations. For all the emission processes, excellent agreement between experiment and theory is generally observed. However, differences in the Ti and Ge peak splittings and energy positions are observed. This provides an important test of the usefulness of theoretical calculations. It also shows that the interpretation of the emission spectra of these materials is best understood as delocalized band states and that excitonic effects are of minor importance.

The analysis of the electronic structures provides increased understanding of the hybridization and chemical bonding of the materials. We find that the Ti-Al bonding in $Ti_3AlC_2$ has less covalent character than the Ti-Si bond of $Ti_3SiC_2$ and Ti-Ge bond of $Ti_3GeC_2$. Emission spectra of Si in $Ti_3SiC_2$, Al in $Ti_3AlC_2$ and Ge in $Ti_3GeC_2$ appear very different from the pure Al, Si and Ge elements indicating strong hybridization between the A-atoms with Ti and C. Changes in the Ti-A bond strength is shown to have direct implications on the Ti-C bonding. Therefore, the A-element substitution and corresponding tuning of the valence electron population from the unfilled states of Al to the isoelectronic states in Si and Ge implies that the unusual material properties of these nanolaminated carbide systems can be custom-made by the choice of A-element.

# 7 Acknowledgements

We would like to thank the staff at MAX-lab for experimental support. This work was supported by the Swedish Research Council, the Göran Gustafsson Foundation, the Swedish Strategic Research Foundation (SSF) Materials Research Programs on Low-Temperature Thin Film Synthesis and the Swedish Agency for Innovation Systems (VINNOVA) Project on Industrialization of MAX Phase Coatings.